\documentclass[10pt,aps,notitlepage,prb,twocolumn,showpacs,preprintnumbers,amsmath,amssymb,floatfix]{revtex4-2}

\usepackage{graphicx}
\usepackage{dsfont}
\usepackage{dcolumn}
\usepackage{bm}
\usepackage{color}
\usepackage{xcolor}
\usepackage{url}
\usepackage{tabularx}
\usepackage{array}
\usepackage{booktabs}
\usepackage{comment}
\usepackage[colorlinks,allcolors=blue]{hyperref}
\usepackage{footnote}
\usepackage{lipsum}
\usepackage[normalem]{ulem}
\usepackage{booktabs}
\usepackage{esint}

\begin{document}
\title{Stability and current-driven dynamics of mixed-topology bimeron clusters}
\author{Philippe Heymes}
\author{Thomas L. Schmidt}
\author{Vladyslav M. Kuchkin}
\email{vladyslav.kuchkin@uni.lu}
\affiliation{Department of Physics and Materials Science, University of Luxembourg, 162A~Avenue de la Faiencerie, L-1511 Luxembourg, Grand Duchy of Luxembourg}

\makeatother

\begin{abstract}
In this work, we study bimerons (in-plane skyrmions) and their clusters composed of states with mixed topological indices. 
These clusters provide flexibility in controlling the skyrmion Hall angle, which depends on the cluster’s total topological charge. 
First, using the simplest such cluster, a bimeron–antibimeron pair, we examine its stability with the geodesic nudged elastic band method and identify three mechanisms -- bimeron separation, interchange, and annihilation -- with comparable energy barriers that determine the pair's overall stability.
This provides an upper bound for the stability of more sophisticated clusters of bimerons and antibimerons. 
Then, we numerically study the clusters' dynamics for currents applied in-plane and perpendicular to the plane, corresponding to the Zhang–Li and Slonczewski mechanisms.
Finally, we analyze the dynamics of a wide variety of bimeron clusters within the Thiele approach and show that their velocities always lie on a specific ellipse whose parameters are fully governed by the model Hamiltonian. 
\end{abstract}

\maketitle

\begin{figure}
\centering
\includegraphics[width=8cm]{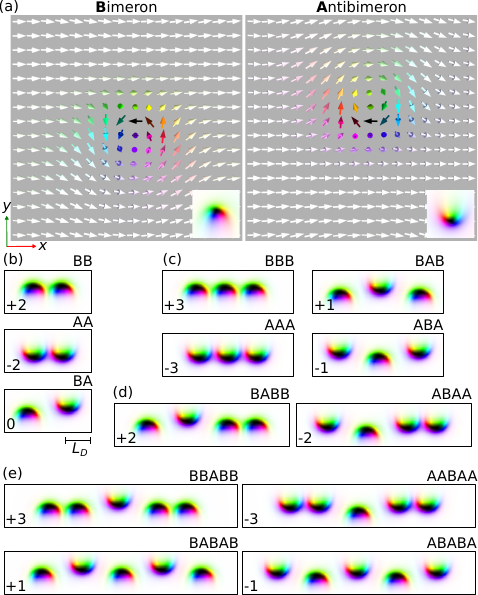}
\caption{~\small \textbf{Bimeron and antibimeron}. Bimerons spin textures are color-coded with HUE given by $\arctan (m_{z}/m_{y}) \in [0, 360^{\circ}]$, while saturation S and value V are given by the $m_{x}$ component of magnetization, i.e., rotated standard color-code~\cite{Savchenko2022}. 
For visualization purposes, we utilize contour plots (the bottom-right insets (a)) instead of the vector fields; this representation is used throughout the paper.
(b)-(e) show stable clusters composed of different numbers of bimerons (B) and antibimeorns (A); values of the topological charge of the cluster are provided in the bottom-left corner.
}
\label{Fig1}
\end{figure}

\noindent
Magnetic bimerons~\cite{Komineas2007, Ezawa2011, Heo2016, Kharkov2017, Leonov2017, Gbel2019} are two-dimensional topological solitons stabilized in an in-plane ferromagnetic background. 
Similar to skyrmions~\cite{Bogdanov_89, Bogdanov_1994}, bimerons are particle-like states that can be classified by the topological charge~\cite{Rybakov2025}: 
\begin{equation}
    Q = \dfrac{1}{4\pi}\int \mathrm{d}x\mathrm{d}y\, \mathbf{m} \cdot [\partial_x \mathbf{m} \times \partial_y \mathbf{m}],\label{Q}
\end{equation}
where unit magnetization vector $\mathbf{m}$ with a fixed point on the spatial infinity, $\mathbf{m}=\mathbf{e}_{x}$ at $|\mathbf{r}|\gg1$.
A primary interest in bimerons is caused by their skyrmion-like stability combined with the possibility to stabilize both the bimeron and its antiparticle, the antibimeron, simultaneously [see Fig.~\ref{Fig1} (a)]. 
Initially, such a particle-antiparticle feature was discussed in the context of magnetic soliton dynamics and the emergent skyrmion Hall angle, which turned out to be proportional to $Q$~\cite{Malozemoff_79}.
However, more recently, quantum simulations of magnetic skyrmions~\cite{Takashima2016, Sotnikov2021, Haller2022, Mazurenko2023, Salvati2024} have revealed that such states might be of interest even statically due to the possibility of forming an entangled state~\cite{Lik2026}, with potential applications for skyrmion qubit operations~\cite{Petrovi2025}.
A similar analysis seems to be applicable for magnetic bimerons as well, although for the purpose of the current work we aim to explore their classical properties, which will be helpful for further quantum case studies.

Bimerons stability is due to the interplay between Heisenberg exchange and Dzyaloshinskii-Moriya (DM) interactions, which can be intrinsic, in materials with broken inversion symmetry~\cite{Yu2010, Yu2010_2, Romming2013, Romming2015, Kzsmrki2015, Nayak2017}, or induced, for instance, in various multilayers containing heavy metals~\cite{Heinze2011, Boulle2016, Yang2018, Chaves2019, Nembach2020}.
In practice, the condition for an in-plane vacuum is easily achieved by a demagnetizing field, leading to the shape anisotropy of a thin film~\cite{hubert2008magnetic}, by an intrinsic magnetocrystalline anisotropy, or by an externally applied magnetic field that forces such magnetization alignment. 
The tilt of the magnetic field to the film plane is known to be responsible for oscillatory asymptotics of skyrmion solutions that were found analytically, and in turn leads to oscillatory interaction potentials~\cite{Kuchkin2020, BartonSinger2023} and the formation of bimerons clusters~\cite{Lin2015, Zhu2026, Yang2026}.
Using different numbers of bimerons and antibimerons, one can construct more sophisticated clusters [Fig.~\ref{Fig1} (b)-(e)], providing us with more flexibility in studying the state with arbitrary topological charge, which defines the dynamical features of the cluster.

Current-driven dynamics of bimerons is an active area of study, with numerous publications on the topic~\cite{Kim2019, Li2020, Murooka2020, Zhang2020, Zarzuela2020, Mukai2024, Vorobyev2026}.
However, more general features such as the geometry of the velocity space~\cite{Kuchkin2021} and the role of soliton symmetry in the dynamics~\cite{Kuchkin2025_1}, reported recently for magnetic skyrmions, are still missing for bimerons and, especially, their clusters.
In this work, we aim to fill this gap and investigate general dynamical properties of bimeron clusters, regardless of the form of DM interactions and the type of electric current applied, as long as they allow cluster stability and motion.

\textit{Micromagnetic model} -- A thin film of a chiral magnet is well described by the micromagnetic Hamiltonian,
\begin{equation}
    \mathcal{E} = l\int \mathrm{d}x\mathrm{d}y\left[\mathcal{A}w_\mathrm{ex} + \mathcal{D}w_\mathrm{DM}+w_{0}\right],\label{Energy_micro}
\end{equation}
where $l$ is the film thickness along which the magnetization remains uniform; $\mathcal{A}$ and $\mathcal{D}$ are strengths of exchange and DM interactions, with energy densities $w_\mathrm{ex}=\left(\nabla m_{x}\right)^{2}+\left(\nabla m_{y}\right)^{2}+\left(\nabla m_{z}\right)^{2}$, and $w_\mathrm{DM}=\mathbf{m}\cdot\nabla\times\mathbf{m}$, respectively, $\nabla = \left(\partial_{x}, \partial_{y}\right)$. 
The potential energy density, $w_{0}=M_\mathrm{s}B(1-m_{x}) - K_{x}m_{x}^{2}-K_{z}m_{z}^{2}$, contains an externally applied field $B>0$ along the $x$-axis, and uniaxial magnetocrystaline anisotropies of easy axis/plane types with strengths $K_{x}$ and $K_{z}$ acting along the $x$- and $z$-axis respectively.
In the limit of thin films, which is the focus of this work, the anisotropy $K_{x}$ and $K_{z}$ can effectively account for the contribution from the demagnetizing field $\mathbf{b}_\mathrm{d}$ created by the sample, i.e. $K_{x}=0, K_{z}<0$ (for a disk) and $K_{x}>0, K_{z}=0$ (for a stripe). 
However, in the simplest form of the Hamiltonian sufficient to stabilize the bimerons, it is enough to have $K_{x}=K_{z}=0$ and apply an external field greater than the so-called skyrmion elliptical instability field, at which bimerons become unstable with respect to stretching deformation~\cite{Bogdanov_1994, BartonSinger2024}. 
The latter can be readily estimated from given material constants, and below we assume that the corresponding parameters ensure bimerons are stable in the uniform vacuum along the applied magnetic field direction.

\begin{figure*}
\centering
\includegraphics[width=18cm]{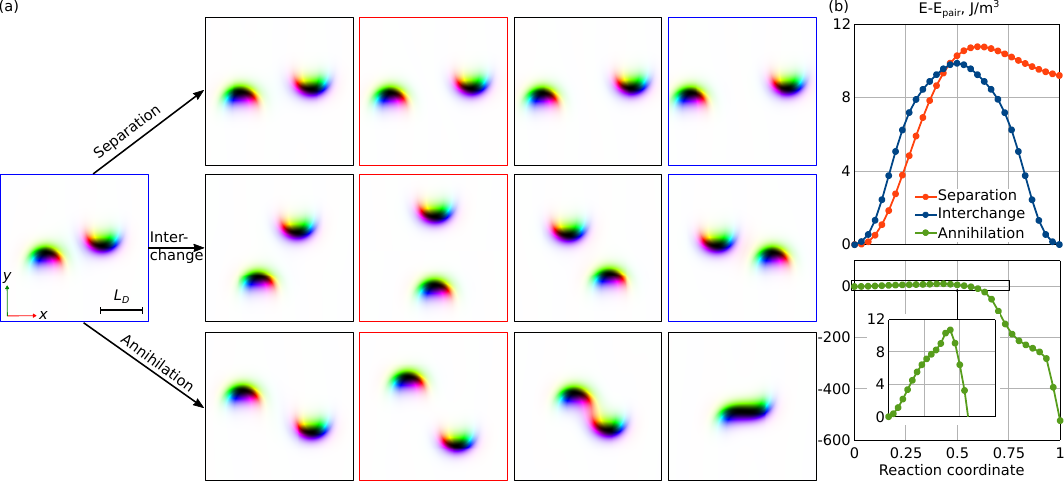}
\caption{~\small \textbf{Bimeron-antibimeron cluster stability}.
(a) Snapshots of the system corresponding to minimum energy path transitions, separation, interchange, and annihilation, respectively, calculated with the geodesic nudged elastic band method. 
Minimum (saddle point) states are highlighted with blue (red) frames.
(b) Energy dependencies for the transitions (a). The rightmost point in panel (b) for the annihilation transition corresponds to the uniform state, $\mathbf{m}(\mathbf{r})=\mathbf{e}_{z}$.
}
\label{Fig2}
\end{figure*}

To describe the dynamics of bimerons, we rely on the LLG equation:
\begin{equation}
    \partial_{t}\mathbf{m} = -\gamma \, \mathbf{m}\times \mathbf{b} + \alpha \, \mathbf{m}\times \partial_{t}\mathbf{m} - \boldsymbol{\tau}, \quad \mathbf{b} = -\dfrac{1}{M_{s}}\dfrac{\delta \mathcal{E}}{\delta \mathbf{m}}, \label{LLG}
\end{equation}
where $\gamma=\gamma_{LL}/(1+\alpha^{2})$ is gyromagnetic ratio, $\alpha$ is Gilbert damping parameter, and $\mathbf{b}$ is an effective field defined by \eqref{Energy_micro}.
An in-plane current of density $\mathbf{j}=(j_{x}, j_{y}, 0)$ can be modeled with Zhang-Li~\cite{Zhang2004} torque ($\boldsymbol{\tau}_{zl}$), while the current $\mathbf{j}=(0, 0, j_{z})$ flowing out-of-plane is described by the Slonczewski~\cite{Slonczewski1996, Xiao2004} torque ($\boldsymbol{\tau}_{sl}$):
\begin{eqnarray}
    &&\boldsymbol{\tau}_{zl} = \left(1-\xi\right)\mathbf{m}\times\left(\mathbf{i}\cdot\nabla\right)\mathbf{m}-\left(\mathbf{i}\cdot\nabla\right)\mathbf{m}, \label{torque_ZL}\\
    &&\boldsymbol{\tau}_{sl} = \gamma\beta\epsilon^{\prime}\mathbf{m}\times\mathbf{p}-\gamma\beta\epsilon\mathbf{m}\times\left(\mathbf{p}\times\mathbf{m}\right), \label{torque_SL}
\end{eqnarray}
where $\xi$ is a non-adiabaticity parameter and a material constant responsible for the efficiency of transferring electron angular momentum to the magnetization, and $\mathbf{i}=-\mu_\mathrm{B}\mu_{0}\mathbf{j}/2e\gamma_{0}\mathcal{M}\left(1+\xi^{2}\right)$ for the in-plane current; $\mathbf{p}=(1,0,0)$ is the fixed layer polarization and $\beta = j_{z}\hbar/M_{s}el$ for the out-of-plane current.
Additionally, $\boldsymbol{\tau}_{sl}$ contains the secondary spin-torque parameter $\epsilon^{\prime}$ and the Slonczewski $\Lambda$ parameter, which characterizes the spacer layer and is used to define $\epsilon$ as
\begin{equation}
    \epsilon = \dfrac{\Lambda^{2}}{\Lambda^{2}+1 + \left(\Lambda^{2}-1\right)\mathbf{m}\cdot\mathbf{p}}. \label{epsilon}
\end{equation}
As can be easily seen, in both cases, the torques \eqref{torque_ZL} and \eqref{torque_SL} do not perturb the vacuum and thus will contribute only to the dynamics of bimerons.
Below, we assume the uniform in time current $\mathbf{j}$, which, however, qualitatively agrees well with the experimental outcomes performed under pulse-type current~\cite{Tang2021, Kern2022, Kern2026}.

In numerical simulations performed in Mumax3 software~\cite{Mumax}, we chose the parameters of FeGe~\cite{Zheng2018}, i.e. $M_{s}=384$ A$/$m, $\mathcal{A}=4.75$ pJ$/$m, $\mathcal{D}= 0.718 $ mJ$/$m$^{2}$ corresponding to the spiral period $L_{D}=4\pi\mathcal{A}/\mathcal{D}\simeq 70$ nm, and external field is $B=0.4 B_{D}$ where $B_{D}=\mathcal{D}^{2}/2\mathcal{A}M_{s}$. Film thickness $l$ is set to $1$ nm.
Additionally we consider anisotropy of strength $K_{x}=0.4 B_{D}M_{s}$.
For GNEB calculations~\cite{Bessarab2015, Bessarab2017}, we rely on Mumax3 extension~\cite{rllg_git, Kuchkin2025_3} and consider a system of size  $6L_{D}\times 6L_{D}$ discretized into $256\times256$ cuboids containing 31 images.
For dynamics simulations, the system of size $12L_{D}\times 6L_{D}$ is discretized into $512\times256$ cuboids.
We fixed $\alpha=0.3$, $\xi=0.6$, $j_{x}=10^{9}$ A$/$m$^{2}$ (for Zhang-Li torque) and $\epsilon^{\prime}=1$, $\Lambda=1$, $j_{z}=10^{8}$ A$/$m$^{2}$ (for Slonczewski torque).
As for the simulations, we always use periodic boundary conditions in the $xy$-plane, so the following formula~\cite{Kuchkin2021, Kuchkin2025_1} for the gravity center $\boldsymbol{\rho}=\left(\rho_{x}, \rho_{y}\right)$ of a bimeron cluster has to be employed:
\begin{eqnarray}
\begin{array}{cc}
     \rho_{x} &= \dfrac{L_x}{2\pi} \tan^{-1} \left( \dfrac{\int N_y \sin(2\pi x / L_x) \, \mathrm{d}x}{\int N_y \cos(2\pi x / L_x) \, \mathrm{d}x} \right) \pm k_x L_x,  \\
     \rho_{y} &= \dfrac{L_y}{2\pi} \tan^{-1} \left( \dfrac{\int N_x \sin(2\pi y / L_y) \, \mathrm{d}y}{\int N_x \cos(2\pi y / L_y) \, \mathrm{d}y} \right) \pm k_y L_y, 
\end{array} 
 \label{mass_center}
\end{eqnarray}
where $N_x \equiv N_x(y) = \int (1 - m_x) \, \mathrm{d}x$ and $N_y \equiv N_y(x) = \int (1 - m_x) \, \mathrm{d}y$, and integer  $k_{x}$, $k_{y}$ are numbers of crossing $x$ and $y$ boundaries.
By taking the time derivative of $\boldsymbol{\rho}$, we obtain the information of the cluster velocity $\mathbf{v}$, which can be further analyzed.

In the rigid motion approximation, $\mathbf{m}\left(\mathbf{r}, t\right)=\mathbf{m}\left(\mathbf{r} - \mathbf{v}t\right)$, the velocity $\mathbf{v}=(v_{x}, v_{y})$ of a spin texture satisfies the Thiele equation~\cite{Thiele_73} followed from the LLG equation~\eqref{LLG}-\eqref{torque_SL}:
\begin{equation}
    Q \, \mathbf{v}\times\mathbf{e}_{z} + \alpha \, \hat{\Gamma} \, \mathbf{v} = \mathbf{f},\label{Thiele_eq}
\end{equation}
where dissipation tensor $\hat{\Gamma}$ has components $\Gamma_{ij}=(4\pi)^{-1}\int\mathrm{d}x\mathrm{d}y \, \partial_i \mathbf{m} \cdot \partial_j \mathbf{m}$.
The force, $\mathbf{f}$, depends on the type of the driving stimuli, and in the case of Zhang-Li torque~\eqref{torque_ZL} one has:
\begin{align}
    \mathbf{f}_{zl}  = Q \,  \mathbf{i} \times \mathbf{e}_{z} + \xi \,\hat{\Gamma} \,\mathbf{i} ,\label{f_zl}
\end{align}
while for the Slonczewski torque~\eqref{torque_SL}: 
\begin{equation}
  \mathbf{f}_{sl}=\dfrac{\gamma \beta}{4\pi} \int  \mathrm{d}x\mathrm{d}y \, \epsilon \left(m_{z} \nabla m_{y} - m_{y} \nabla m_{z}\right).\label{f_sl}
\end{equation}
Note that $\epsilon$ depends on $\mathbf{m}$ according to \eqref{epsilon}, which has to be taken into account while calculating the integral \eqref{f_sl}.

\begin{figure*}
\centering
\includegraphics[width=18cm]{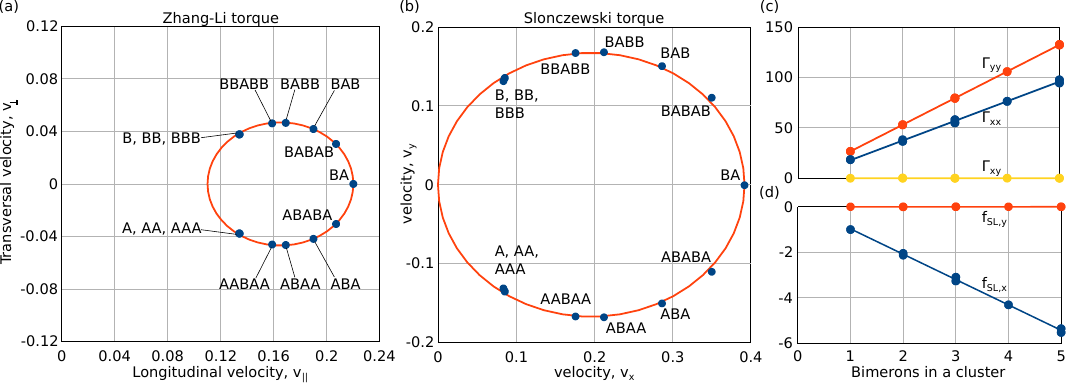}
\caption{~\small \textbf{Results of LLG simulations}.
(a) and (b) velocity space for bimeron clusters [Fig.~\ref{Fig1}(b)-(e)] under Zhang-Li and Slonczewski torques, respectively.
Points correspond to data extracted from LLG simulations; red ellipses are given by Eqs.~(\ref{ZL_sol_ellipse}) and (\ref{SL_sol_ellipse}) obtained from the Thiele equation analysis. 
(c) and (d) show dependencies of dissipation tensor components and force \eqref{f_sl} on the total number of bimerons in a cluster.
}
\label{Fig3}
\end{figure*}

\textit{Bimeron-antibimeron cluster stability} -- Clusters composed of topologically identical bimerons do not demonstrate very distinct dynamics from a single bimeron simply because the Thiele equation contains integral quantities which linearly scale with the number of bimerons.
The situation is different when one uses bimerons of different topological charges for the cluster.
Here we analyze the stability of the simplest cluster of such type composed of a bimeron and an antibimeron with total charge $Q=0$.
Relying on the GNEB method for minimum energy path (MEP) calculations~\cite{mumax3_scripts}, we consider three transitions: separation, interchange, and annihilation [see Fig.~\ref{Fig2}(a)] determining the cluster stability. These transitions have comparable energy barriers, as shown in MEPs (b).
The last transition is interesting from a topological perspective and reflects the possibility of bimeron and antibimeron to annihilate each other without the appearance of singularities. 
Right before transforming into the saturated state, one can notice droplet-like soliton nucleation~\cite{Kuchkin2022}, which, however, is not stable at the parameters we use.
As we are interested in not too high current strengths, which will preserve the number of bimeron particles, this transition is not relevant for further dynamics simulations and is shown here only for generality.
The transitions corresponding to bimerons separation and interchange are characterized by slightly different energy barriers, $E_{s}$ and $E_{i}$, which indicates one may expect certain asymmetry also in their dynamics, e.g., dependence on the direction of the motion. 
In particular, enforcing a bimeron pair to move left and right, achieved simply by flipping the current direction, will result in different critical currents at which separation and interchange instability occurs. 
Therefore, denoting these critical currents $J_{+}$ and $J_{-}$, we can assume $J_{+}-J_{-}\sim E_{s}-E_{i} \neq 0$,   which can be further experimentally tested.

\textit{Dynamics of various bimeron clusters} -- By numerically solving the LLG equation~\cite{mumax3_scripts}, we study the dynamics of the bimeron clusters shown in Fig.~\ref{Fig1}(b)-(e). The applied current strength is kept low enough not to modify the clusters' shape, but high enough to induce their motion.
Extracting information about each cluster's gravity center \eqref{mass_center}, we are able to obtain the cluster velocity, which is plotted in Fig.~\ref{Fig3}(a) and (b), for the Zhang-Li and Slonczewski torques, respectively.
The velocities are obtained in the regime when clusters move with a constant velocity, i.e., in the steady motion regime.
All data points with sufficient accuracy lie on a certain ellipse; below, we analyze this in more detail.
Velocities of clusters composed of identical particles (only bimerons or only antibimerons) coincide with high accuracy.
After exchanging all particles with antiparticles in a cluster, its transversal, or $y$-component, velocity changes sign due to the so-called skyrmion Hall angle.
The bimeron-antibimeron (BA) cluster moves without a deflection angle (i.e., $v_{y}=0$), and we expect the same for other types of topologically trivial clusters. 

Although the underlying physical mechanisms of the Zhang-Li and Slonczewski torques differ, comparing the cluster dynamics allows us to deduce only quantitative distinctions in the resulting clusters' velocities. 
This might indicate that the type of the driving force plays a secondary role in magnetic soliton dynamics, while the primary role belongs to the underlying dynamical equation, i.e., the LLG equation.
This is even more evident from the Thiele equation, whose form is dictated solely by the LLG equation.

\textit{Analysis of the velocity space} -- Analyzing spin textures of the bimeron and the antibemeron [Fig.~\ref{Fig1}](a), we can write the symmetry operation that maps one onto the other as follows:
\begin{eqnarray}
    \left(m^{B}_{x},m^{B}_{y},m^{B}_{z}\right)(x,y) = \left(m^{A}_{x},m^{A}_{z},m^{A}_{y}\right)(y, x), \label{BA_symmetry}
\end{eqnarray}
where superscript $B$ and $A$ in magnetization components $m_{i}$ stays for bimeron and antibimeron, respectively.
This symmetry property results in a relationship between dissipation tensor components:
\begin{equation}
    \Gamma_{ij}\left(\mathbf{m}^{B}\right) = \left(2\delta_{ij}-1\right)\Gamma_{ji}\left(\mathbf{m}^{A}\right), \, i,j \in \{x,y\} \label{dis_tensor_AB}.
\end{equation}
This, in turn, indicates that the total number of particles, $N_{B}+N_{A}$, in a cluster matters for $\Gamma_{ij}$, rather than their type.
Moreover, as each $\Gamma_{ij}$ is an integral quantity, it scales almost linearly with this number, i.e., one can write 
 $\Gamma_{ij}=\left(N_{B}+N_{A}\right)\Gamma_{ij}^{1}$,
where $\Gamma_{ij}^{1}$ is dissipation tensor components calculated for one (anti-)bimeron. 
This is also supported by numerical simulations [Fig.~\ref{Fig3}(c)].
At the same time, the topological charge depends on the type of particles and can be computed as $Q=N_{B}-N_{A}$.

In the case of Zhang-Li torque, using the force \eqref{f_zl}, we write the solution of the Thiele equation~\eqref{Thiele_eq} as follows~\cite{ku_thesis}:
\begin{equation}
    \!\!\!\!v_\parallel = i - i \frac{\left( \alpha \Gamma + Q \Gamma_{xy} \right) (\alpha-\xi)}{Q^2 + \alpha^2 \Gamma},
v_\perp = i \frac{(\alpha-\xi ) Q}{Q^2 + \alpha^2 \Gamma} \Gamma_{xx},\label{ZL_sol}
\end{equation}
where $\Gamma = \Gamma_{xx}\Gamma_{yy}-\Gamma_{xy}^{2}$, $i = \sqrt{i_{x}^{2}+i_{y}^{2}}$, $v_{\parallel}$ and $v_{\perp}$ are, respectively, longitudinal and transversal velocities relative to the current direction.
Assuming that $|\Gamma_{xy}| \ll \Gamma_{ii}$, $i=x,y$, as follows from numerics [see Fig.~\ref{Fig3}(c)], we can simplify \eqref{ZL_sol} as follows:
\begin{eqnarray}
&&\!\!\!\!v_\parallel = i\dfrac{\xi+\alpha}{2\alpha}+i\dfrac{\xi-\alpha}{2\alpha}\dfrac{1 - \nu^{2}}{1 + \nu^{2}},v_\perp = i\dfrac{\xi-\alpha}{2\alpha} \dfrac{2 \nu}{1 + \nu^{2}} \sqrt{\dfrac{\Gamma_{xx}^{1}}{\Gamma_{yy}^{1}}},\nonumber\\
&&\!\!\!\!\nu = -\dfrac{1}{\alpha\sqrt{\Gamma_{xx}^{1}\Gamma_{yy}^{1}}}\dfrac{N_{B}-N_{A}}{N_{B}+N_{A}}.\label{ZL_sol_vxvy}
\end{eqnarray}%
Denoting $\nu=\tan \psi/2$, we can recover the Weierstrass substitution and write the following equation bounding $v_\parallel$ and $v_{\perp}$:
\begin{equation}
    \left(v_\parallel - i\dfrac{\xi+\alpha}{2\alpha}\right)^{2} + v_{\perp}^{2}\dfrac{\Gamma_{yy}^{1}}{\Gamma_{xx}^{1}} = \left(i\dfrac{\xi-\alpha}{2\alpha}\right)^{2}.\label{ZL_sol_ellipse}
\end{equation}
Expression \eqref{ZL_sol_ellipse} defines an ellipse with aspect ratio $\sqrt{\Gamma_{yy}^{1}/\Gamma_{xx}^{1}}$ shifted by $i\left(\xi+\alpha\right)/2\alpha$ along the current direction.

For the Slonczewski torque and the force~\eqref{f_sl}, we get:
\begin{align}
\begin{array}{cc}
     v_x &= \dfrac{\alpha \Gamma_{yy}\,f_{x}   -(\alpha \Gamma_{xy}-Q)\,f_{y}}{Q^2 + \alpha^2 \Gamma},   \\
     v_y &= \dfrac{(\alpha \Gamma_{xy}+Q)\,f_{x}-\alpha \Gamma_{xx}\,f_{y}}{Q^2 + \alpha^2 \Gamma}. 
\end{array}
    \label{SL_sol}
\end{align}
Additionally to $|\Gamma_{xy}| \ll \Gamma_{ii}$, $i=x,y$, we assume $|f_{y}|\ll |f_{x}|$ and $f_{x}=\left(N_{B}+N_{A}\right)f_{x}^{1}$, as follows from numerics [see Fig.~\ref{Fig3}(d)], and simplify \eqref{SL_sol} as follows:
\begin{eqnarray}
    v_{x}=\dfrac{f_{x}^{1}}{\alpha \Gamma_{xx}^{1}}\dfrac{1}{1 + \nu^{2}}, \,\,v_{y} =-\dfrac{f_{x}^{1}}{\alpha \sqrt{\Gamma_{xx}^{1}\Gamma_{yy}^{1}}}\dfrac{\nu}{1 + \nu^{2}},
\end{eqnarray}
which, after the Weierstrass substitution, provides us with the following ellipse equation:
\begin{equation}
    \left(v_{x} - \dfrac{f_{x}^{1}}{2\alpha \Gamma_{xx}^{1}}\right)^{2} + v_{y}^{2}\dfrac{\Gamma_{yy}^{1}}{\Gamma_{xx}^{1}}= \left(\dfrac{f_{x}^{1}}{2\alpha \Gamma_{xx}^{1}}\right)^{2}.\label{SL_sol_ellipse}
\end{equation}
The aspect ratio of ellipse \eqref{SL_sol_ellipse} is again given by $\sqrt{\Gamma_{yy}^{1}/\Gamma_{xx}^{1}}$.
Thus, the underlying Hamiltonian plays a role in the static characteristics of a bimeron, $\Gamma_{xx}$ and $\Gamma_{yy}$, which define the geometry of the velocity space.

Geometrically, the only qualitative distinction between ellipses \eqref{ZL_sol_ellipse} and \eqref{SL_sol_ellipse} is that the latter goes through origin (i.e., it contains the point $(v_{x}, v_{y})=(0,0,)$), while for the former the leftmost point on the velocity space is $(v_{\parallel}, v_{\perp}) = (i,0)$ (for $\xi>\alpha$) or $(v_{\parallel}, v_{\perp}) = (i\xi/\alpha,0)$ (for $\xi<\alpha$).
The exact position of the cluster velocity on the ellipse depends on the particle type and is preserved across different arrangements as long as the cluster is constructed with the same number of bimerons and antibimerons. 

Taking into account previously reported helicity-dependent dynamics 
 of skyrmions under Slonczewski torque~\cite{Kim2018, Msiska2022}, which allows control of the Hall angle via DM interaction type (Bloch-Neel), we expect the same effect for bimerons, manifesting as an ellipse's primary-axis dependence on helicity. 
This is evident from the form of the force \eqref{f_sl}, which depends on the magnetic texture helicity, while other terms in the Thiele equation \eqref{Thiele_eq} ($\Gamma_{ij}$ and $Q$) are helicity-invariant.
At the same time, dynamics under the Zhang-Li torque is free from this effect and thus is physically distinct from the Slonczewski torque case.

\section*{Conclusions}
In this work, we studied clusters of magnetic bimerons in a thin layer of chiral magnets.
Performing simulations with the geodesic nudged elastic band method, we identified three mechanisms: separation, interchange, and annihilation, which define the total stability of a cluster.
We then studied cluster dynamics induced by in-plane and out-of-plane currents, modeled through Zhang-Li and Slonczewski torques.
A main outcome of these numerical results was the geometry of the velocity space, having the shape of an ellipse, to which the velocities of all bimeron clusters belong. 
Analysis based on the Thiele equation solutions supports these observations and provides analytical equations for the ellipses.
The parameters of the ellipse are defined by the Hamiltonian (aspect ratio) and dynamical characteristics (current, damping, torque-related material constants).

\section*{Acknowledgments}
V.M.K acknowledges the financial support from the European Union’s Horizon Europe research and innovation programme under the Marie Sk{\l}odowska-Curie grant agreement No.~101203692 (QUANTHOPF). We acknowledge the financial support from the Luxembourg National Research Fund under Grants C22/MS/17415246/DeQuSky and AFR/23/17951349.

\section*{Author contributions}
P.H. and V.M.K. performed the dynamics simulations and Thiele analysis.
V.M.K. performed minimum energy path calculations and wrote the first draft. 
All authors interpreted the results and contributed to writing and revising the manuscript.

\section*{Data Availability} 
The code and raw data sets are published on Zenodo~\cite{mumax3_scripts}.

\newpage

\bibliographystyle{apsrev4-2}
\bibliography{main}

@article{Kim2018,
  title = {Asymmetric skyrmion Hall effect in systems with a hybrid Dzyaloshinskii-Moriya interaction},
  author = {Kim, Kyoung-Whan and Moon, Kyoung-Woong and Kerber, Nico and Nothhelfer, Jonas and Everschor-Sitte, Karin},
  journal = {Phys. Rev. B},
  volume = {97},
  issue = {22},
  pages = {224427},
  numpages = {10},
  year = {2018},
  month = {Jun},
  publisher = {American Physical Society},
  url = {https://link.aps.org/doi/10.1103/PhysRevB.97.224427}
}

@article{Msiska2022,
  title = {Nonzero Skyrmion Hall Effect in Topologically Trivial Structures},
  volume = {17},
  ISSN = {2331-7019},
  url = {http://dx.doi.org/10.1103/PhysRevApplied.17.064015},
  number = {6},
  journal = {Physical Review Applied},
  publisher = {American Physical Society (APS)},
  author = {Msiska,  Robin and Rodrigues,  Davi R. and Leliaert,  Jonathan and Everschor-Sitte,  Karin},
  year = {2022},
  month = June 
}

@article{BartonSinger2024,
  title = {Modelling the Elliptical Instability of Magnetic Skyrmions},
  volume = {4},
  ISSN = {2673-8724},
  url = {http://dx.doi.org/10.3390/magnetism4040020},
  number = {4},
  journal = {Magnetism},
  publisher = {MDPI AG},
  author = {Barton-Singer,  Bruno},
  year = {2024},
  month = Sept,
  pages = {305–321}
}

@article{Leonov2017,
  title = {Asymmetric isolated skyrmions in polar magnets with easy-plane anisotropy},
  volume = {96},
  ISSN = {2469-9969},
  url = {http://dx.doi.org/10.1103/PhysRevB.96.014423},
  number = {1},
  journal = {Physical Review B},
  publisher = {American Physical Society (APS)},
  author = {Leonov,  A. O. and Kézsmárki,  I.},
  year = {2017},
  month = July 
}

@article{Romming2013,
  title = {Writing and Deleting Single Magnetic Skyrmions},
  volume = {341},
  ISSN = {1095-9203},
  url = {http://dx.doi.org/10.1126/science.1240573},
  number = {6146},
  journal = {Science},
  publisher = {American Association for the Advancement of Science (AAAS)},
  author = {Romming,  Niklas and Hanneken,  Christian and Menzel,  Matthias and Bickel,  Jessica E. and Wolter,  Boris and von Bergmann,  Kirsten and Kubetzka,  André and Wiesendanger,  Roland},
  year = {2013},
  month = Aug,
  pages = {636–639}
}

@article{Nembach2020,
  title = {Correlation between Dzyaloshinskii-Moriya interaction and orbital angular momentum at an oxide-ferromagnet interface},
  volume = {101},
  ISSN = {2469-9969},
  url = {http://dx.doi.org/10.1103/PhysRevB.101.020409},
  number = {2},
  journal = {Physical Review B},
  publisher = {American Physical Society (APS)},
  author = {Nembach,  Hans T. and Jué,  Emilie and Evarts,  Eric R. and Shaw,  Justin M.},
  year = {2020},
  month = Jan 
}

@book{ku_thesis,
  url = {https://publications.rwth-aachen.de/record/855848},
  author = {Kuchkin,  Vladyslav},
  title = {Diversity of chiral magnetic solitons},
  journal = {Dissertation},
  volume = {RWTH Aachen University},
  pages = {2022},
  publisher = {RWTH Aachen University},
  year = {2022}
}

@article{Kern2022,
  title = {Deterministic Generation and Guided Motion of Magnetic Skyrmions by Focused He<sup>+</sup>-Ion Irradiation},
  volume = {22},
  ISSN = {1530-6992},
  url = {http://dx.doi.org/10.1021/acs.nanolett.2c00670},
  number = {10},
  journal = {Nano Letters},
  publisher = {American Chemical Society (ACS)},
  author = {Kern,  Lisa-Marie and Pfau,  Bastian and Deinhart,  Victor and Schneider,  Michael and Klose,  Christopher and Gerlinger,  Kathinka and Wittrock,  Steffen and Engel,  Dieter and Will,  Ingo and G\"{u}nther,  Christian M. and Liefferink,  Rein and Mentink,  Johan H. and Wintz,  Sebastian and Weigand,  Markus and Huang,  Meng-Jie and Battistelli,  Riccardo and Metternich,  Daniel and B\"{u}ttner,  Felix and H\"{o}flich,  Katja and Eisebitt,  Stefan},
  year = {2022},
  month = May,
  pages = {4028–4035}
}

@article{Kern2026,
  title = {Emergent Chaos‐Like Dynamics of Spin–Orbit‐Torque‐Driven Magnetic Transitions},
  volume = {22},
  ISSN = {1613-6829},
  url = {http://dx.doi.org/10.1002/smll.73778},
  number = {37},
  journal = {Small},
  publisher = {Wiley},
  author = {Kern,  L.‐M. and Litzius,  K. and Deinhart,  V. and Schneider,  M. and Klose,  C. and Gerlinger,  K. and Battistelli,  R. and Metternich,  D. and Engel,  D. and G\"{u}nther,  C. M. and Huang,  M.‐J. and H\"{o}flich,  K. and B\"{u}ttner,  F. and Eisebitt,  S. and Pfau,  B.},
  year = {2026},
  month = May 
}

@article{Tang2021,
  title = {Magnetic skyrmion bundles and their current-driven dynamics},
  volume = {16},
  ISSN = {1748-3395},
  url = {http://dx.doi.org/10.1038/s41565-021-00954-9},
  number = {10},
  journal = {Nature Nanotechnology},
  publisher = {Springer Science and Business Media LLC},
  author = {Tang,  Jin and Wu,  Yaodong and Wang,  Weiwei and Kong,  Lingyao and Lv,  Boyao and Wei,  Wensen and Zang,  Jiadong and Tian,  Mingliang and Du,  Haifeng},
  year = {2021},
  month = Aug,
  pages = {1086–1091}
}

@article{Chaves2019,
  title = {Oxidation dependence of the {Dzyaloshinskii}-{Moriya} interaction in {Pt}/{Co}/{MO}$_x$ trilayers ({M} = {Al} or {Gd})},
  author = {Chaves, Dayane de Souza and Ajejas, Fernando and K\v{r}i\v{z}\'{a}kov\'{a}, Viola and Vogel, Jan and Pizzini, Stefania},
  journal = {Physical Review B},
  volume = {99},
  number = {14},
  pages = {144404},
  year = {2019},
  month = apr,
  publisher = {American Physical Society},
  url = {https://doi.org/10.1103/PhysRevB.99.144404},
  issn = {2469-9950}
}

@article{Yang2018,
  title = {Controlling Dzyaloshinskii-Moriya Interaction via Chirality Dependent Atomic-Layer Stacking,  Insulator Capping and Electric Field},
  volume = {8},
  ISSN = {2045-2322},
  url = {http://dx.doi.org/10.1038/s41598-018-30063-y},
  number = {1},
  journal = {Scientific Reports},
  publisher = {Springer Science and Business Media LLC},
  author = {Yang,  Hongxin and Boulle,  Olivier and Cros,  Vincent and Fert,  Albert and Chshiev,  Mairbek},
  year = {2018},
  month = Aug 
}

@article{Boulle2016,
  title = {Room-temperature chiral magnetic skyrmions in ultrathin magnetic nanostructures},
  volume = {11},
  ISSN = {1748-3395},
  url = {http://dx.doi.org/10.1038/nnano.2015.315},
  number = {5},
  journal = {Nature Nanotechnology},
  publisher = {Springer Science and Business Media LLC},
  author = {Boulle,  Olivier and Vogel,  Jan and Yang,  Hongxin and Pizzini,  Stefania and de Souza Chaves,  Dayane and Locatelli,  Andrea and Menteş,  Tevfik Onur and Sala,  Alessandro and Buda-Prejbeanu,  Liliana D. and Klein,  Olivier and Belmeguenai,  Mohamed and Roussigné,  Yves and Stashkevich,  Andrey and Chérif,  Salim Mourad and Aballe,  Lucia and Foerster,  Michael and Chshiev,  Mairbek and Auffret,  Stéphane and Miron,  Ioan Mihai and Gaudin,  Gilles},
  year = {2016},
  month = Jan,
  pages = {449–454}
}

@article{Heinze2011,
  title = {Spontaneous atomic-scale magnetic skyrmion lattice in two dimensions},
  volume = {7},
  ISSN = {1745-2481},
  url = {http://dx.doi.org/10.1038/nphys2045},
  number = {9},
  journal = {Nature Physics},
  publisher = {Springer Science and Business Media LLC},
  author = {Heinze,  Stefan and von Bergmann,  Kirsten and Menzel,  Matthias and Brede,  Jens and Kubetzka,  André and Wiesendanger,  Roland and Bihlmayer,  Gustav and Bl\"{u}gel,  Stefan},
  year = {2011},
  month = July,
  pages = {713–718}
}

@article{Nayak2017,
  title = {Magnetic antiskyrmions above room temperature in tetragonal Heusler materials},
  volume = {548},
  ISSN = {1476-4687},
  url = {http://dx.doi.org/10.1038/nature23466},
  number = {7669},
  journal = {Nature},
  publisher = {Springer Science and Business Media LLC},
  author = {Nayak,  Ajaya K. and Kumar,  Vivek and Ma,  Tianping and Werner,  Peter and Pippel,  Eckhard and Sahoo,  Roshnee and Damay,  Franoise and R\"{o}ßler,  Ulrich K. and Felser,  Claudia and Parkin,  Stuart S. P.},
  year = {2017},
  month = Aug,
  pages = {561–566}
}

@article{Yu2010,
  title = {Near room-temperature formation of a skyrmion crystal in thin-films of the helimagnet FeGe},
  volume = {10},
  ISSN = {1476-4660},
  url = {http://dx.doi.org/10.1038/nmat2916},
  number = {2},
  journal = {Nature Materials},
  publisher = {Springer Science and Business Media LLC},
  author = {Yu,  X. Z. and Kanazawa,  N. and Onose,  Y. and Kimoto,  K. and Zhang,  W. Z. and Ishiwata,  S. and Matsui,  Y. and Tokura,  Y.},
  year = {2010},
  month = Dec,
  pages = {106–109}
}

@article{Yu2010_2,
  title = {Real-space observation of a two-dimensional skyrmion crystal},
  volume = {465},
  ISSN = {1476-4687},
  url = {http://dx.doi.org/10.1038/nature09124},
  number = {7300},
  journal = {Nature},
  publisher = {Springer Science and Business Media LLC},
  author = {Yu,  X. Z. and Onose,  Y. and Kanazawa,  N. and Park,  J. H. and Han,  J. H. and Matsui,  Y. and Nagaosa,  N. and Tokura,  Y.},
  year = {2010},
  month = June,
  pages = {901–904}
}

@article{Kuchkin2022,
  title = {Thermal generation of droplet soliton in chiral magnet},
  volume = {105},
  ISSN = {2469-9969},
  url = {http://dx.doi.org/10.1103/PhysRevB.105.184403},
  number = {18},
  journal = {Physical Review B},
  publisher = {American Physical Society (APS)},
  author = {Kuchkin,  Vladyslav M. and Bessarab,  Pavel F. and Kiselev,  Nikolai S.},
  year = {2022},
  month = May 
}

@article{Romming2015,
  title = {Field-Dependent Size and Shape of Single Magnetic Skyrmions},
  volume = {114},
  ISSN = {1079-7114},
  url = {http://dx.doi.org/10.1103/PhysRevLett.114.177203},
  number = {17},
  journal = {Physical Review Letters},
  publisher = {American Physical Society (APS)},
  author = {Romming,  Niklas and Kubetzka,  André and Hanneken,  Christian and von Bergmann,  Kirsten and Wiesendanger,  Roland},
  year = {2015},
  month = May 
}

@article{Kzsmrki2015,
  title = {Néel-type skyrmion lattice with confined orientation in the polar magnetic semiconductor GaV4S8},
  volume = {14},
  ISSN = {1476-4660},
  url = {http://dx.doi.org/10.1038/nmat4402},
  number = {11},
  journal = {Nature Materials},
  publisher = {Springer Science and Business Media LLC},
  author = {Kézsmárki,  I. and Bordács,  S. and Milde,  P. and Neuber,  E. and Eng,  L. M. and White,  J. S. and Rønnow,  H. M. and Dewhurst,  C. D. and Mochizuki,  M. and Yanai,  K. and Nakamura,  H. and Ehlers,  D. and Tsurkan,  V. and Loidl,  A.},
  year = {2015},
  month = Sept,
  pages = {1116–1122}
}

@article{Mukai2024,
  title = {“Polymerization” of Bimerons in Quasi-Two-Dimensional Chiral Magnets with Easy-Plane Anisotropy},
  volume = {14},
  ISSN = {2079-4991},
  url = {http://dx.doi.org/10.3390/nano14060504},
  number = {6},
  journal = {Nanomaterials},
  publisher = {MDPI AG},
  author = {Mukai,  Natsuki and Leonov,  Andrey O.},
  year = {2024},
  month = Mar,
  pages = {504}
}

@article{Zhang2020,
  title = {Static and dynamic properties of bimerons in a frustrated ferromagnetic monolayer},
  volume = {101},
  ISSN = {2469-9969},
  url = {http://dx.doi.org/10.1103/PhysRevB.101.144435},
  number = {14},
  journal = {Physical Review B},
  publisher = {American Physical Society (APS)},
  author = {Zhang,  Xichao and Xia,  Jing and Shen,  Laichuan and Ezawa,  Motohiko and Tretiakov,  Oleg A. and Zhao,  Guoping and Liu,  Xiaoxi and Zhou,  Yan},
  year = {2020},
  month = Apr 
}

@Article{Bessarab2015,
  author    = {Bessarab, Pavel F. and Uzdin, Valery M. and Jónsson, Hannes},
  journal   = {Computer Physics Communications},
  title     = {Method for finding mechanism and activation energy of magnetic transitions, applied to skyrmion and antivortex annihilation},
  year      = {2015},
  issn      = {0010-4655},
  month     = nov,
  pages     = {335–347},
  volume    = {196},
  publisher = {Elsevier BV},
  url       = {http://dx.doi.org/10.1016/j.cpc.2015.07.001},
}

@Article{Bessarab2017,
  author    = {Bessarab, P. F.},
  journal   = {Physical Review B},
  title     = {Comment on “{Path} to collapse for an isolated {Néel} skyrmion”},
  year      = {2017},
  issn      = {2469-9969},
  month     = apr,
  number    = {13},
  pages     = {136401},
  volume    = {95},
  publisher = {American Physical Society (APS)},
  url       = {http://dx.doi.org/10.1103/PhysRevB.95.136401},
}

@article{Zarzuela2020,
  title = {Stability and dynamics of in-plane skyrmions in collinear ferromagnets},
  volume = {101},
  ISSN = {2469-9969},
  url = {http://dx.doi.org/10.1103/PhysRevB.101.054405},
  number = {5},
  journal = {Physical Review B},
  publisher = {American Physical Society (APS)},
  author = {Zarzuela,  Ricardo and Bharadwaj,  Venkata Krishna and Kim,  Kyoung-Whan and Sinova,  Jairo and Everschor-Sitte,  Karin},
  year = {2020},
  month = Feb 
}

@article{Murooka2020,
  title = {Current-induced shuttlecock-like movement of non-axisymmetric chiral skyrmions},
  volume = {10},
  ISSN = {2045-2322},
  url = {http://dx.doi.org/10.1038/s41598-019-56791-3},
  number = {1},
  journal = {Scientific Reports},
  publisher = {Springer Science and Business Media LLC},
  author = {Murooka,  Remi and Leonov,  Andrey O. and Inoue,  Katsuya and Ohe,  Jun-ichiro},
  year = {2020},
  month = Jan 
}

@article{Lin2015,
  title = {Skyrmion fractionalization and merons in chiral magnets with easy-plane anisotropy},
  volume = {91},
  ISSN = {1550-235X},
  url = {http://dx.doi.org/10.1103/PhysRevB.91.224407},
  number = {22},
  journal = {Physical Review B},
  publisher = {American Physical Society (APS)},
  author = {Lin,  Shi-Zeng and Saxena,  Avadh and Batista,  Cristian D.},
  year = {2015},
  month = June 
}

@article{Zhang2004,
  title = {Roles of Nonequilibrium Conduction Electrons on the Magnetization Dynamics of Ferromagnets},
  volume = {93},
  ISSN = {1079-7114},
  url = {http://dx.doi.org/10.1103/PhysRevLett.93.127204},
  number = {12},
  journal = {Physical Review Letters},
  publisher = {American Physical Society (APS)},
  author = {Zhang,  S. and Li,  Z.},
  year = {2004},
  month = Sept 
}

@article{Kim2019,
  title = {Dynamics of bimeron skyrmions in easy-plane magnets induced by a spin supercurrent},
  volume = {99},
  ISSN = {2469-9969},
  url = {http://dx.doi.org/10.1103/PhysRevB.99.224406},
  number = {22},
  journal = {Physical Review B},
  publisher = {American Physical Society (APS)},
  author = {Kim,  Se Kwon},
  year = {2019},
  month = June 
}

@article{Li2020,
  title = {Bimeron clusters in chiral antiferromagnets},
  volume = {6},
  ISSN = {2057-3960},
  url = {http://dx.doi.org/10.1038/s41524-020-00435-y},
  number = {1},
  journal = {npj Computational Materials},
  publisher = {Springer Science and Business Media LLC},
  author = {Li,  Xiaoguang and Shen,  Laichuan and Bai,  Yuhao and Wang,  Junlin and Zhang,  Xichao and Xia,  Jing and Ezawa,  Motohiko and Tretiakov,  Oleg A. and Xu,  Xiaohong and Mruczkiewicz,  Michal and Krawczyk,  Maciej and Xu,  Yongbing and Evans,  Richard F. L. and Chantrell,  Roy W. and Zhou,  Yan},
  year = {2020},
  month = Nov 
}

@article{Zhu2026,
  title = {Light-induced bimerons in a chiral magnet},
  volume = {17},
  ISSN = {2041-1723},
  url = {http://dx.doi.org/10.1038/s41467-026-71291-5},
  number = {1},
  journal = {Nature Communications},
  publisher = {Springer Science and Business Media LLC},
  author = {Zhu,  Kaixin and Rybakov,  Filipp N. and Wang,  Zhan and Gao,  Wenli and Sun,  Shuaishuai and Wang,  Wentao and Li,  Jun and Tian,  Huanfang and Eriksson,  Olle and Yang,  Huaixin and Zhang,  Ying and Kiselev,  Nikolai S. and Li,  Zian and Li,  Jianqi},
  year = {2026},
  month = Apr 
}

@article{Vorobyev2026,
  title = {Asymmetric antibimerons: Statics and dynamics},
  volume = {1},
  ISSN = {3070-2240},
  url = {http://dx.doi.org/10.1103/kdtb-q3ly},
  journal = {APS Open Science},
  publisher = {American Physical Society (APS)},
  author = {Vorobyev,  Pavel A. and Kurebayashi,  Daichi and Tretiakov,  Oleg A.},
  year = {2026},
  month = July 
}

@article{Mumax,
  title = {The design and verification of MuMax3},
  volume = {4},
  ISSN = {2158-3226},
  url = {http://dx.doi.org/10.1063/1.4899186},
  number = {10},
  journal = {AIP Advances},
  publisher = {AIP Publishing},
  author = {Vansteenkiste,  Arne and Leliaert,  Jonathan and Dvornik,  Mykola and Helsen,  Mathias and Garcia-Sanchez,  Felipe and Van Waeyenberge,  Bartel},
  year = {2014},
  month = oct 
}

@article{Thiele_73,
  title = {Steady-State Motion of Magnetic Domains},
  volume = {30},
  ISSN = {0031-9007},
  url = {http://dx.doi.org/10.1103/PhysRevLett.30.230},
  number = {6},
  journal = {Physical Review Letters},
  publisher = {American Physical Society (APS)},
  author = {Thiele,  A. A.},
  year = {1973},
  month = feb,
  pages = {230–233}
}

@article{Savchenko2022,
  title = {Chiral standing spin waves in skyrmion lattice},
  volume = {10},
  ISSN = {2166-532X},
  url = {http://dx.doi.org/10.1063/5.0097651},
  number = {7},
  journal = {APL Materials},
  publisher = {AIP Publishing},
  author = {Savchenko,  Andrii S. and Kuchkin,  Vladyslav M. and Rybakov,  Filipp N. and Bl\"{u}gel,  Stefan and Kiselev,  Nikolai S.},
  year = {2022},
  month = July 
}

@article{Kuchkin2025_3,
  title = {Coexisting mechanisms of thermally driven magnetization reversal in shakti spin ice systems},
  volume = {7},
  ISSN = {2643-1564},
  url = {http://dx.doi.org/10.1103/pr2y-dfbd},
  number = {3},
  journal = {Physical Review Research},
  publisher = {American Physical Society (APS)},
  author = {Kuchkin,  Vladyslav M. and Arnalds,  Unnar B. and Jónsson,  Hannes and Bessarab,  Pavel F.},
  year = {2025},
  month = Sept 
}

@article{Haller2022,
  title = {Quantum skyrmion lattices in Heisenberg ferromagnets},
  volume = {4},
  ISSN = {2643-1564},
  url = {http://dx.doi.org/10.1103/PhysRevResearch.4.043113},
  number = {4},
  journal = {Physical Review Research},
  publisher = {American Physical Society (APS)},
  author = {Haller,  Andreas and Groenendijk,  Solofo and Habibi,  Alireza and Michels,  Andreas and Schmidt,  Thomas L.},
  year = {2022},
  month = Nov 
}

@article{Salvati2024,
  title = {Stability of a quantum skyrmion: Projective measurements and the quantum Zeno effect},
  volume = {109},
  ISSN = {2469-9969},
  url = {http://dx.doi.org/10.1103/PhysRevB.109.064409},
  number = {6},
  journal = {Physical Review B},
  publisher = {American Physical Society (APS)},
  author = {Salvati,  Fabio and Katsnelson,  Mikhail I. and Bagrov,  Andrey A. and Westerhout,  Tom},
  year = {2024},
  month = Feb 
}

@book{hubert2008magnetic,
  title={Magnetic domains: the analysis of magnetic microstructures},
  author={Hubert, Alex and Sch{\"a}fer, Rudolf},
  year={2008},
  publisher={Springer Science \& Business Media}
}

@article{Komineas2007,
  title = {Rotating Vortex Dipoles in Ferromagnets},
  volume = {99},
  ISSN = {1079-7114},
  url = {http://dx.doi.org/10.1103/PhysRevLett.99.117202},
  number = {11},
  journal = {Physical Review Letters},
  publisher = {American Physical Society (APS)},
  author = {Komineas,  S.},
  year = {2007},
  month = Sept 
}

@article{Heo2016,
  title = {Switching of chiral magnetic skyrmions by picosecond magnetic field pulses via transient topological states},
  volume = {6},
  ISSN = {2045-2322},
  url = {http://dx.doi.org/10.1038/srep27146},
  number = {1},
  journal = {Scientific Reports},
  publisher = {Springer Science and Business Media LLC},
  author = {Heo,  Changhoon and Kiselev,  Nikolai S. and Nandy,  Ashis Kumar and Bl\"{u}gel,  Stefan and Rasing,  Theo},
  year = {2016},
  month = June 
}

@article{Kharkov2017,
  title = {Bound States of Skyrmions and Merons near the Lifshitz Point},
  volume = {119},
  ISSN = {1079-7114},
  url = {http://dx.doi.org/10.1103/PhysRevLett.119.207201},
  number = {20},
  journal = {Physical Review Letters},
  publisher = {American Physical Society (APS)},
  author = {Kharkov,  Y. A. and Sushkov,  O. P. and Mostovoy,  M.},
  year = {2017},
  month = Nov 
}

@article{Gbel2019,
  title = {Magnetic bimerons as skyrmion analogues in in-plane magnets},
  volume = {99},
  ISSN = {2469-9969},
  url = {http://dx.doi.org/10.1103/PhysRevB.99.060407},
  number = {6},
  journal = {Physical Review B},
  publisher = {American Physical Society (APS)},
  author = {G\"{o}bel,  B\"{o}rge and Mook,  Alexander and Henk,  J\"{u}rgen and Mertig,  Ingrid and Tretiakov,  Oleg A.},
  year = {2019},
  month = Feb 
}

@article{Ezawa2011,
  title = {Compact merons and skyrmions in thin chiral magnetic films},
  volume = {83},
  ISSN = {1550-235X},
  url = {http://dx.doi.org/10.1103/PhysRevB.83.100408},
  number = {10},
  journal = {Physical Review B},
  publisher = {American Physical Society (APS)},
  author = {Ezawa,  Motohiko},
  year = {2011},
  month = Mar 
}

@article{Bogdanov_89,
  author  = {A. N. Bogdanov and D. A. Yablonskii},
  journal = {Sov. Phys. JETP},
  number  = {101},
  title   = {Thermodynamically stable ``vortices'' in magnetically ordered crystals. The mixed state of magnets},
  url     = {http://www.jetp.ac.ru/cgi-bin/e/index/e/68/1/p101?a=list},
  volume  = {68},
  year    = {1989}
}

@article{Bogdanov_1994,
  author    = {A. Bogdanov and A. Hubert},
  journal   = {physica status solidi (b)},
  month     = dec,
  number    = {2},
  pages     = {527--543},
  publisher = {Wiley},
  title     = {The Properties of Isolated Magnetic Vortices},
  url       = {https://doi.org/10.1002/pssb.2221860223},
  volume    = {186},
  year      = {1994}
}

@article{Kuchkin2025_1,
  title = {Symmetry-governed dynamics of magnetic skyrmions under field pulses},
  volume = {8},
  ISSN = {2399-3650},
  url = {http://dx.doi.org/10.1038/s42005-024-01913-1},
  number = {1},
  journal = {Communications Physics},
  publisher = {Springer Science and Business Media LLC},
  author = {Kuchkin,  Vladyslav M. and Barton-Singer,  Bruno and Bessarab,  Pavel F. and Kiselev,  Nikolai S.},
  year = {2025},
  month = Jan 
}

@article{Kuchkin2021,
  title = {Geometry and symmetry in skyrmion dynamics},
  volume = {104},
  ISSN = {2469-9969},
  url = {http://dx.doi.org/10.1103/PhysRevB.104.165116},
  number = {16},
  journal = {Physical Review B},
  publisher = {American Physical Society (APS)},
  author = {Kuchkin,  Vladyslav M. and Chichay,  Ksenia and Barton-Singer,  Bruno and Rybakov,  Filipp N. and Bl\"{u}gel,  Stefan and Schroers,  Bernd J. and Kiselev,  Nikolai S.},
  year = {2021},
  month = Oct 
}

@article{BartonSinger2023,
  title = {Stability and asymptotic interactions of chiral  magnetic skyrmions in a tilted magnetic field},
  volume = {15},
  ISSN = {2542-4653},
  url = {http://dx.doi.org/10.21468/SciPostPhys.15.1.011},
  number = {1},
  journal = {SciPost Physics},
  publisher = {Stichting SciPost},
  author = {Barton-Singer,  Bruno and Schroers,  Bernd},
  year = {2023},
  month = July 
}

@article{Kuchkin2020,
  title = {Turning a chiral skyrmion inside out},
  volume = {101},
  ISSN = {2469-9969},
  url = {http://dx.doi.org/10.1103/PhysRevB.101.064408},
  number = {6},
  journal = {Physical Review B},
  publisher = {American Physical Society (APS)},
  author = {Kuchkin,  Vladyslav M. and Kiselev,  Nikolai S.},
  year = {2020},
  month = Feb 
}

@article{Zheng2018,
  title = {Experimental observation of chiral magnetic bobbers in B20-type FeGe},
  volume = {13},
  ISSN = {1748-3395},
  url = {http://dx.doi.org/10.1038/s41565-018-0093-3},
  number = {6},
  journal = {Nature Nanotechnology},
  publisher = {Springer Science and Business Media LLC},
  author = {Zheng,  Fengshan and Rybakov,  Filipp N. and Borisov,  Aleksandr B. and Song,  Dongsheng and Wang,  Shasha and Li,  Zi-An and Du,  Haifeng and Kiselev,  Nikolai S. and Caron,  Jan and Kovács,  András and Tian,  Mingliang and Zhang,  Yuheng and Bl\"{u}gel,  Stefan and Dunin-Borkowski,  Rafal E.},
  year = {2018},
  month = apr,
  pages = {451–455}
}

@book{Malozemoff_79,
  title={Magnetic domain walls in bubble materials: advances in materials and device research},
  author={Malozemoff, AP and Slonczewski, John C},
  volume={1},
  year={1979},
  publisher={Academic press}
}

@article{Petrovi2025,
  title = {Colloquium
                    : Quantum properties and functionalities of magnetic skyrmions},
  volume = {97},
  ISSN = {1539-0756},
  url = {http://dx.doi.org/10.1103/RevModPhys.97.031001},
  number = {3},
  journal = {Reviews of Modern Physics},
  publisher = {American Physical Society (APS)},
  author = {Petrović,  Alexander P. and Psaroudaki,  Christina and Fischer,  Peter and Garst,  Markus and Panagopoulos,  Christos},
  year = {2025},
  month = July 
}

@article{Mazurenko2023,
  title = {Estimating Patterns of Classical and Quantum Skyrmion States},
  volume = {92},
  ISSN = {1347-4073},
  url = {http://dx.doi.org/10.7566/JPSJ.92.081004},
  number = {8},
  journal = {Journal of the Physical Society of Japan},
  publisher = {Physical Society of Japan},
  author = {Mazurenko,  Vladimir V. and Iakovlev,  Ilia A. and Sotnikov,  Oleg M. and Katsnelson,  Mikhail I.},
  year = {2023},
  month = Aug 
}

@article{Takashima2016,
  title = {Quantum skyrmions in two-dimensional chiral magnets},
  volume = {94},
  ISSN = {2469-9969},
  url = {http://dx.doi.org/10.1103/PhysRevB.94.134415},
  number = {13},
  journal = {Physical Review B},
  publisher = {American Physical Society (APS)},
  author = {Takashima,  Rina and Ishizuka,  Hiroaki and Balents,  Leon},
  year = {2016},
  month = Oct 
}

@article{Sotnikov2021,
  title = {Probing the topology of the quantum analog of a classical skyrmion},
  volume = {103},
  ISSN = {2469-9969},
  url = {http://dx.doi.org/10.1103/PhysRevB.103.L060404},
  number = {6},
  journal = {Physical Review B},
  publisher = {American Physical Society (APS)},
  author = {Sotnikov,  O. M. and Mazurenko,  V. V. and Colbois,  J. and Mila,  F. and Katsnelson,  M. I. and Stepanov,  E. A.},
  year = {2021},
  month = Feb 
}

@article{Lik2026,
  title = {Skyrmionic Schr\"{o}dinger cat states in monoaxial chiral magnets},
  volume = {113},
  ISSN = {2469-9969},
  url = {http://dx.doi.org/10.1103/842q-ghyt},
  number = {17},
  journal = {Physical Review B},
  publisher = {American Physical Society (APS)},
  author = {Liscak, Stefan and Haller, Andreas and Michels, Andreas and Schmidt, Thomas L. and Kuchkin, Vladyslav M.},
  year = {2026},
  month = May 
}

\end{document}